\documentclass[12pt]{article}

\usepackage{amsfonts,amssymb,bm,cite,amsmath}
\usepackage[dvipdfmx]{graphicx}
\usepackage{array, booktabs}
\usepackage{braket}
\usepackage[title]{appendix}

\newcommand {\beq}{\begin{equation}}
\newcommand {\eeq}{\end{equation}}
\newcommand {\beqa}{\begin{eqnarray}}
\newcommand {\eeqa}{\end{eqnarray}}

\allowdisplaybreaks

\begin{document}

\setlength{\oddsidemargin}{0cm}
\setlength{\baselineskip}{7mm}

\begin{titlepage}
\renewcommand{\thefootnote}{\fnsymbol{footnote}}
\begin{normalsize}
\begin{flushright}
\begin{tabular}{l}
November 2020
\end{tabular}
\end{flushright}
\end{normalsize}

~~\\

\vspace*{0cm}
    \begin{Large}
       \begin{center}
         {Renormalization group and diffusion equation }
       \end{center}
    \end{Large}
\vspace{1cm}

\begin{center}
           Masami M{\sc atsumoto}\footnote
            {
e-mail address : 
matsumoto.masami.16@shizuoka.ac.jp},
           Gota T{\sc anaka}\footnote
            {
e-mail address : 
tanaka.gota.14@shizuoka.ac.jp}
                   and
           Asato T{\sc suchiya}\footnote
           {
e-mail address : 
tsuchiya.asato@shizuoka.ac.jp}\\
      \vspace{1cm}

 {\it Department of Physics, Shizuoka University}\\
                {\it 836 Ohya, Suruga-ku, Shizuoka 422-8529, Japan}\\
         \vspace{0.3cm}     
  {\it Graduate School of Science and Technology, Shizuoka University}\\
               {\it 3-5-1 Johoku, Naka-ku, Hamamatsu 432-8011, Japan}

\end{center}

\vspace{3cm}

\begin{abstract}
\noindent
We study the relationship between the renormalization group 
and the diffusion equation. We consider the
exact renormalization group equation 
for a scalar field that includes an arbitrary cutoff function and
an arbitrary quadratic seed action. As a generalization of the result obtained by Sonoda and Suzuki, 
we find that the correlation functions of diffused fields with respect to the bare action
agree with those of bare fields with respect to the effective action,
where the diffused field obeys a generalized diffusion equation determined by the cutoff function
and the seed action and  agrees 
with the bare field at the initial time.
\end{abstract}
\vfill
\end{titlepage}
\vfil\eject

\setcounter{footnote}{0}

\section{Introduction}
\setcounter{equation}{0}
\renewcommand{\thefootnote}{\arabic{footnote}}

It is recognized that the diffusion equation is connected with the
renormalization group, since diffusion can be regarded as 
a continuum analog of block-spin transformation or coarse-graining.
Indeed, the solution to the diffusion equation in $d$ dimensions
\begin{align}
\partial_{\tau}\varphi(\tau,x)=\partial_i^2 \varphi(\tau,x) 
\label{diffusion equation}
\end{align}
with an initial condition $\varphi(0,x)=\phi(x)$ is given by
\begin{align}
\varphi(\tau,x)=\int d^dx'\ K(x,x',\tau)\phi(x') \ ,
\label{real space solution to diffusion equation}
\end{align}
where $K(x,x',\tau)$ is the heat kernel
\begin{align}
K(x,x',\tau) = \frac{1}{(4\pi \tau)^{d/2}}\exp\left[-\frac{(x-x')^2}{4\tau}\right] \ .
\label{heat kernel}
\end{align}
Here, $\varphi(\tau,x)$ can be viewed as a coarse-grained field obtained by 
smearing  $\phi$ in a ball centered at $x$ with radius $\sqrt{\tau}$.
On the other hand,
the gradient flow equation in gauge theories, which is regarded as a generalized 
diffusion equation respecting gauge symmetry, has recently been used to
renormalize composite operators and so on \cite{Narayanan:2006rf,Luscher:2009eq,Luscher:2010iy}.

Thus it is natural to expect that there is relationship between
(generalized) diffusion equations or gradient flow equations and renormalization group equations.
Indeed, the relationship has been studied in 
\cite{Kagimura:2015via,Capponi:2015ucc,Aoki:2016ohw,Aoki:2017bru,Makino:2018rys,Abe:2018zdc,
Carosso:2019tan,Sonoda:2019ibh}.
In particular, it was shown in \cite{Sonoda:2019ibh} 
that the correlation functions of the diffused fields (\ref{real space solution to diffusion equation}) 
with respect to the bare action
agree with those of bare fields with respect to the effective action, where
the effective action obeys the exact renormalization group (ERG) equation with a cutoff function and a seed action.
This implies that correlation functions of the composite operators consisting of
the diffused fields (\ref{real space solution to diffusion equation}) are finite 
so that the diffused fields can be used to renormalize the composite operators.

In this paper, we study a generalization of  the result in \cite{Sonoda:2019ibh}.
We consider the ERG equation for a  scalar field with an arbitrary cutoff function and
an arbitrary quadratic seed action.
We show that the correlation functions of diffused fields with respect to the bare action
agree with those of bare fields with respect to the effective action obeying the above
ERG equation, 
where the diffused field obeys a generalized diffusion equation determined
by the cutoff function and the seed action and agrees 
with the bare field at the initial $\tau$.

We also perform the $\epsilon$ expansion using the derivative expansion 
as a check of validity of the ERG equation.
We reproduce the well-known scaling dimensions of operators around the Wilson-Fisher fixed point.


Throughout this paper, except for section 5, we work in the momentum space, and introduce the notation
\begin{align}
\int_p \equiv \int \frac{d^dp}{(2\pi)^d} \ .
\end{align}
Note that the Fourier transforms 
of (\ref{diffusion equation}),  (\ref{real space solution to diffusion equation})
and (\ref{heat kernel})
are, respectively, 
\begin{align}
(\partial_{\tau} + p^2) \varphi(\tau,p) = 0  \ ,
\label{diffusion equation in momentum space}
\end{align}
\begin{align}
\varphi(\tau,p) = K(p,\tau)  \phi(p) \ ,
\label{diffused field in momentum space}
\end{align}
\begin{align}
K(p,\tau) = e^{-\tau p^2}  \ .
\end{align}

This paper is organized as follows.
In section 2, we give our statement on the relationship between the ERG equation
and a generalized diffusion equation.
In section 3, we prove our statement using functional differential equations for generating functionals of 
the correlation functions.
In section 4, we provide 
another proof of our statement by solving the ERG equation
using a functional integration kernel.
In section 5, we perform the $\epsilon$ expansion using the derivative expansion.
We reproduce the well-known scaling dimensions of operators around the Wilson-Fisher fixed point.
Section 6 is devoted to the conclusion and a discussion. 
We clarify the reason for restricting ourselves to seed actions that are quadratic in $\phi$.

\section{Relation between renormalization group and diffusion equation}
\setcounter{equation}{0}
The ERG equation
is a functional differential equation that describes nonperturbatively how the effective action
$S_{\Lambda}$ at the energy scale $\Lambda$ changes when 
$\Lambda$ is decreased to $\Lambda - d\Lambda$.
The one for a scalar field in $d$ dimensions is specified by
a cutoff function $\dot{C}_{\Lambda}(p^2)$ and a seed action $\hat{S}_{\Lambda}$,
and takes the form \cite{Latorre:2000qc,Arnone:2002yh,Arnone:2005fb,Morris:1999px}
\begin{align}
	-\Lambda \frac{\partial}{\partial \Lambda} e^{-S_\Lambda[\phi]} 
	= &-\frac{1}{2} \int_p  \dot{C}_\Lambda(p^2)\frac{\delta^2}{\delta \phi(p)\delta \phi(-p)}  
e^{-S_\Lambda[\phi]}
- \int_p \dot{C}_\Lambda(p^2) \frac{\delta}{\delta \phi(p)} \left( \frac{\delta \hat{S}_\Lambda[\phi]}{\delta \phi(-p)} e^{-S_\Lambda[\phi]}	 \right)  \ .
\label{exact renormalization group equation}
\end{align}
$\dot{C}_{\Lambda}$ is quasi-local and incorporates UV regularization, while the seed action
$\hat{S}_{\Lambda}$ is a functional of $\phi$ that has derivative expansion.
Here we consider an arbitrary cutoff function $\dot{C}_{\Lambda}(p^2)$ and an arbitrary quadratic seed action
that satisfy the above conditions.
The coarse-graining procedure is fixed by the cutoff function and the seed action.
We parametrize the seed action as
\begin{align}
\hat{S}_{\Lambda}[\phi]=-\frac{1}{2}\int_p \dot{C}^{-1}_{\Lambda}(p^2) \chi_{\Lambda}(p^2)
\phi(p)\phi(-p) \ ,
\label{quadratic seed action}
\end{align}
where $\chi_{\Lambda}$ is an arbitrary regular function.
The initial condition for (\ref{exact renormalization group equation}) is given at a bare cutoff scale
$\Lambda_0$, and $S_{\Lambda_0}=S_{\Lambda=\Lambda_0}$ is a bare action.

We define the vacuum expectation value with respect to the effective action $S_{\Lambda}$ as
\begin{align}
\langle \cdots \rangle_{\Lambda}
=\frac{1}{Z_{\Lambda}} \int {\cal D}\phi \  \cdots \  e^{-S_{\Lambda}[\phi]}
\end{align}
with
\begin{align}
Z_{\Lambda}=\int {\cal D}\phi \  e^{-S_{\Lambda}[\phi]}  \ .
\end{align}

We show the following relation on the correlation functions
\begin{align}
\langle \prod_{a=1}^n \phi(p_a) \rangle_{\Lambda}^c
=\langle \prod_{a=1}^n \varphi(\tau,p_a) \rangle_{\Lambda_0}^c  
+\delta_{n,2} \ (2\pi)^d\delta^d(p_1+p_2)  \ r(\Lambda,p_1^2) \ ,
\label{relation}
\end{align}
where $c$ stands for the connected part, and
\begin{align}
\tau = \frac{1}{\Lambda^2} -\frac{1}{\Lambda_0^2} \ .
\label{tau}
\end{align}
$\varphi(\tau,p)$ is a solution to the differential equation
\begin{align}
\partial_{\tau} \varphi(\tau,p) = \frac{1}{2}\Lambda^2 \dot{C}_{\Lambda}(p^2)
\frac{\delta \hat{S}_{\Lambda}[\varphi]}{\delta \varphi(\tau,-p)} = -\frac{1}{2}\Lambda^2\chi_{\Lambda}(p^2)
\varphi(\tau,p) \ ,
\label{general diffusion equation}
\end{align}
or, equivalently,
\begin{align}
-\Lambda \frac{\partial}{\partial \Lambda} \varphi(\tau,p)
= \dot{C}_{\Lambda}(p^2)\frac{\delta \hat{S}_{\Lambda}[\varphi]}{\delta \varphi(\tau,-p)} 
=-\chi_{\Lambda}(p^2)\varphi(\tau,p) \ 
\label{general diffusion equation 2}
\end{align}
with the initial condition
\begin{align}
\varphi(0,p) = \phi(p) \ .
\label{initial condition for varphi}
\end{align}
$r(\Lambda,p^2)$ obeys
the differential equation
\begin{align}
\Lambda \frac{\partial r(\Lambda,p^2)}{\partial \Lambda}=\dot{C}_{\Lambda}(p^2)+
2\chi_{\Lambda}(p^2)r(\Lambda,p^2) \ ,
\label{differential equation for r}
\end{align}
and satisfies the initial condition
\begin{align}
r(\Lambda_0,p^2)=0 \ .
\label{initial condition for r}
\end{align}
(\ref{general diffusion equation}) or (\ref{general diffusion equation 2})  can be viewed as
a generalized diffusion equation.
(\ref{relation}) gives a relation between the ERG equation and the 
generalized diffusion equation.

For later convenience, we solve (\ref{differential equation for r}) and (\ref{general diffusion equation 2}).
The solution to (\ref{differential equation for r}) with the initial condition
(\ref{initial condition for r}) is given by 
\begin{align}
r(\Lambda,p^2)= A^{-1}_{\Lambda,\Lambda_0}(p^2)B^{-2}_{\Lambda,\Lambda_0}(p^2)
\label{explicit form of r}
\end{align}
with
\begin{align}
&A_{\Lambda,\Lambda_0}(p^2)=\left[ -\int_{\Lambda}^{\Lambda_0} \frac{d\Lambda'}{\Lambda'}
\dot{C}_{\Lambda'}(p^2)B^2_{\Lambda',\Lambda_0}(p^2)\right]^{-1}   \ ,   \label{A} \\
&B_{\Lambda,\Lambda_0}(p^2)=\exp \left[\int_{\Lambda}^{\Lambda_0}\frac{d\Lambda'}{\Lambda'}
\chi_{\Lambda'}(p^2)\right] \ .
\label{B}
\end{align}
The solution to (\ref{general diffusion equation 2}) with the initial condition (\ref{initial condition for varphi})
is  given by
\begin{align}
\varphi(\tau,p) = B^{-1}_{\Lambda,\Lambda_0}(p^2) \phi(p) \ .
\label{explicit form of varphi}
\end{align}

We will  give a proof of (\ref{relation}) in the next section. Before closing this section, we give
an example of $\dot{C}_{\Lambda}$ and $\chi_{\Lambda}(p^2)$:
\begin{align}
\dot{C}_{\Lambda}(p^2) =-\frac{2}{\Lambda^2}K_{\Lambda}(p^2)+\frac{\zeta}{p^2}K_{\Lambda}(p^2)
(1-K_{\Lambda}(p^2))
\label{example of cdot}
\end{align}
with
\begin{align}
K_{\Lambda}(p^2) = e^{-p^2/\Lambda^2} \ ,
\end{align}
and
\begin{align}
\chi_{\Lambda}(p^2)
=2\frac{p^2}{\Lambda^2}  - \frac{\eta}{2} \ .
\label{example of chi}
\end{align}
Setting $\zeta = \eta$ in (\ref{example of cdot}) yields
an ERG equation studied in \cite{Sonoda:2019ibh,Sonoda:2015bla}.
We rescale $\varphi$ as
\begin{align}
\varphi'(\tau,p) = \left( \frac{\Lambda}{\Lambda_0}\right)^{\frac{\eta}{2}} \varphi(\tau,p) \ .
\end{align}
Then, (\ref{relation}) reduces to
\begin{align}
\langle \prod_{a=1}^n \phi(p_a) \rangle_{\Lambda}^c
=&\left(\frac{\Lambda_0}{\Lambda}\right)^{\frac{n\eta}{2}}  
\langle \prod_{a=1}^n \varphi'(\tau,p_a) \rangle_{\Lambda_0}^c   \nonumber\\
&+\delta_{n,2} \ (2\pi)^d\delta^d(p_1+p_2)  \frac{K_{\Lambda}^2(p^2)}{p^2}
\left\{ \frac{1-K_{\Lambda}(p^2)}{K_{\Lambda}(p^2)}-\left(\frac{\Lambda_0}{\Lambda}\right)^{\eta}
\frac{1-K_{\Lambda_0}(p^2)}{K_{\Lambda_0}(p^2)} \right\}   \ .
\label{S-S relation}
\end{align}
We see from (\ref{general diffusion equation}) that
$\varphi'$ satisfies (\ref{diffusion equation in momentum space}) and is given by
(\ref{diffused field in momentum space}).
(\ref{S-S relation}) is nothing but the relation obtained in \cite{Sonoda:2019ibh}.

\section{Proof of the relation (\ref{relation})}
\setcounter{equation}{0}
In this section we give a proof of the relation (\ref{relation}).
For this purpose, we define two functionals of $J(p)$:
\begin{align}
U[J(\cdot),\Lambda]  &= \ln\int {\cal D}\phi \  e^{i \int_p J(p)\phi(-p)} \ e^{-S_{\Lambda}[\phi]} \ , 
\label{U} \\
V[J(\cdot),\Lambda] &= V'[J(\cdot),\Lambda] + R[J(\cdot),\Lambda]
\end{align}
with 
\begin{align}
&V'[J(\cdot),\Lambda]=\ln \int {\cal D}\phi \  e^{i \int_p J(p)\varphi(\tau,-p)} \ e^{-S_{\Lambda_0}[\phi]} \ ,
\\
&R[J(\cdot),\Lambda]=\frac{i^2}{2}\int_p r(\Lambda,p^2) J(p)J(-p) \ .
\label{R}
\end{align}
$U$ is the generating functional for the connected correlation functions of $\phi$ with respect to $S_{\Lambda}$,
while $V'$ is the one for the connected correlation functions of $\varphi$ with respect to $S_{\Lambda_0}$.
$U$ and $V$ agree at $\Lambda=\Lambda_0$
because $\varphi$ and $r(\Lambda,p^2)$ satisfy (\ref{initial condition for varphi})
and (\ref{initial condition for r}), respectively:
\begin{align}
U[J(\cdot),\Lambda_0] = V[J(\cdot),\Lambda_0] \ .
\label{initial condition for U and V}
\end{align}

In the following,
we will show that $U$ and $V$ satisfy the same functional differential equation,
which is the first order in the  $\Lambda$ derivative.
First, we calculate
\begin{align}
-\Lambda \frac{\partial}{\partial \Lambda} U[J(\cdot),\Lambda]
\end{align}
as follows:
\begin{align}
& =\frac{1}{e^U}\int {\cal D}\phi \  e^{i \int_p J(p)\phi(-p)} \ 
\left(-\Lambda \frac{\partial}{\partial \Lambda}
e^{-S_{\Lambda}[\phi]} \right)  \nonumber\\
&= \frac{1}{e^U}\int {\cal D}\phi \  e^{i \int_p J(p)\phi(-p)}   \  \left(  
 -\frac{1}{2} \int_p \dot{C}_{\Lambda}(p^2) \frac{\delta^2}{\delta \phi(p) \delta \phi(-p)} 
e^{-S_{\Lambda}[\phi]}   
 - \int_p \dot{C}_{\Lambda}(p^2) \frac{\delta}{\delta \phi(p)} 
\left(\frac{\delta \hat{S}_{\Lambda}[\phi]}{\delta \phi(-p)} 
e^{-S_{\Lambda}[\phi]} \right)  \right)   \nonumber\\
& = -\frac{i^2}{2} \int_p \dot{C}_{\Lambda}(p^2) J(p)J(-p) 
+\frac{i}{e^U} \int {\cal D}\phi \  e^{i \int_p J(p)\phi(-p)} \int_p \dot{C}_{\Lambda}(p^2) J(p)
\frac{\delta \hat{S}_{\Lambda}[\phi]}{\delta \phi(p)} e^{-S_{\Lambda}[\phi]}  \nonumber\\
& =  -\frac{i^2}{2} \int_p \dot{C}_{\Lambda}(p^2) J(p)J(-p) 
+\frac{i}{e^U} \int_p \dot{C}_{\Lambda}(p^2) J(p)
\frac{\delta \hat{S}_{\Lambda}[\phi]}{\delta \phi(p)} \left[ \frac{1}{i}\frac{\delta}{\delta J(\cdot)}    \right]  e^U
\nonumber\\
& = -\frac{i^2}{2} \int_p \dot{C}_{\Lambda}(p^2) J(p)J(-p) 
-\int_p \chi_{\Lambda}(p^2) J(p) \frac{\delta U}{\delta J(p)} \ .
\label{calculation for U}
\end{align}
Here we have used (\ref{exact renormalization group equation}) in the second equality and performed
partial path integrations in the third equality.
In the fourth, 
$\frac{\delta \hat{S}_{\Lambda}[\phi]}{\delta \phi(p)} \left[ \frac{1}{i}\frac{\delta}{\delta J(\cdot)}    \right] $
is obtained by replacing $\phi(p')$ in a functional $\frac{\delta \hat{S}_{\Lambda}[\phi]}{\delta \phi(p)}$
with $\frac{1}{i}\frac{\delta}{\delta J(-p')}$.
In the fifth, we used the explicit form of $\hat{S}_{\Lambda}$ (\ref{quadratic seed action}).

Next, we calculate
\begin{align}
-\Lambda \frac{\partial}{\partial \Lambda} V[J(\cdot),\Lambda]
\end{align}
as follows:
\newpage
\begin{align}
&=-\frac{i^2}{2} \int_p \Lambda\frac{\partial r(\Lambda,p^2)}{\partial \Lambda} J(p)J(-p) 
+\frac{i}{e^{V'}} \int {\cal D}\phi \  e^{i \int_p J(p)\varphi(\tau,-p)} 
\int_p J(p) \left(-\Lambda\frac{\partial}{\partial \Lambda} \varphi(\tau,-p) \right)
e^{-S_{\Lambda_0}[\phi]}
\nonumber\\
&=-\frac{i^2}{2} \int_p (\dot{C}_{\Lambda}(p^2)+2\chi_{\Lambda}(p^2)r(\Lambda,p^2) ) J(p)J(-p)
\nonumber\\
&\;\;\;\;+\frac{i}{e^{V-R}} \int {\cal D}\phi \  e^{i \int_p J(p)\varphi(\tau,-p)} 
\int_p J(p) \dot{C}_{\Lambda}(p^2)\frac{\delta \hat{S}_{\Lambda}[\varphi]}{\delta\varphi(\tau,p)}
e^{-S_{\Lambda_0[\phi]}}
\nonumber\\
& = -\frac{i^2}{2} \int_p (\dot{C}_{\Lambda}(p^2)+2\chi_{\Lambda}(p^2)r(\Lambda,p^2) ) J(p)J(-p) 
+\frac{i}{e^{V-R}} \int_p  \dot{C}_{\Lambda}(p^2) J(p)
\frac{\delta \hat{S_{\Lambda}}[\phi]}{\delta\phi(p)}\left[\frac{1}{i}\frac{\delta}{\delta J(\cdot)}\right] e^{V-R}
\nonumber\\
& = -\frac{i^2}{2} \int_p (\dot{C}_{\Lambda}(p^2)+2\chi_{\Lambda}(p^2)r(\Lambda,p^2) ) J(p)J(-p) 
-\int_p  \chi_{\Lambda}(p^2) J(p) \frac{\delta}{\delta J(p)} (V-R)
\nonumber\\
& = -\frac{i^2}{2} \int_p \dot{C}_{\Lambda}(p^2) J(p)J(-p) 
-\int_p  \chi_{\Lambda}(p^2) J(p) \frac{\delta V}{\delta J(p)} 
\label{calculation for V}
\end{align}
Here we have used (\ref{general diffusion equation 2}) 
and (\ref{differential equation for r}) in the second equality, 
(\ref{quadratic seed action}) in the fourth equality and
(\ref{R}) in the last equality.

We see from (\ref{calculation for U}) and (\ref{calculation for V}) 
that $U$ and $V$ satisfy the same functional differential equation,
which is the first order in the  $\Lambda$ derivative.
We therefore conclude with (\ref{initial condition for U and V}) that
\begin{align}
U[J(\cdot),\Lambda] = V[J(\cdot),\Lambda] \ .
\label{U=V}
\end{align}

%
Finally, we calculate the left-hand side of the relations (\ref{relation}):
\begin{align}
\langle \prod_{a=1}^n \phi(p_a) \rangle_{\Lambda}^c 
&= \left. \frac{1}{i^n}\frac{\delta^n}{\delta J(-p_1) \cdots \delta J(-p_n)}U[J(\cdot),\Lambda] \right|_{J=0} \nonumber\\
&= \left. \frac{1}{i^n}\frac{\delta^n}{\delta J(-p_1) \cdots \delta J(-p_n)}V[J(\cdot),\Lambda] \right|_{J=0}
\nonumber\\
&= \left. \frac{1}{i^n}\frac{\delta^n}{\delta J(-p_1) \cdots \delta J(-p_n)}V'[J(\cdot),\Lambda] \right|_{J=0}
+\delta_{n,2}(2\pi)^d\delta^d(p_1+p_2) r(\Lambda,p_1^2)
\nonumber\\
&= \langle \prod_{a=1}^n \varphi(\tau,p_a) \rangle_{\Lambda_0}^c  +\delta_{n,2}(2\pi)^d\delta^d(p_1+p_2) r(\Lambda,p_1^2) \ .
\end{align}
Thus, we have completed the proof of (\ref{relation}).

\section{Functional integration kernel}
\setcounter{equation}{0}
In this section we show (\ref{relation}) by solving
(\ref{exact renormalization group equation}) in terms of a functional integration kernel,
which is a functional analog of the heat kernel (\ref{heat kernel}).
We seek for a functional integration kernel ${\cal K}[\phi,\phi',\Lambda,\Lambda_0]$ that is a solution to
the ERG equation (\ref{exact renormalization group equation}) satisfying
the initial condition
\begin{align}
{\cal K}[\phi,\phi',\Lambda_0,\Lambda_0]=\Delta [\phi-\phi'] \ ,
\label{initial condition for kernel}
\end{align}
where $\Delta$ is the delta functional.
It is easy to show that ${\cal K}$ is given by
\begin{align}
&{\cal K}[\phi,\phi',\Lambda,\Lambda_0]  \nonumber\\
&=\mbox{det}[A^{\frac{1}{2}}_{\Lambda,\Lambda_0}B_{\Lambda,\Lambda_0}]
\exp\left[-\frac{1}{2}\int_p A_{\Lambda,\Lambda_0}(p^2)
(B_{\Lambda,\Lambda_0}(p^2)\phi(p)-\phi'(p))(B_{\Lambda,\Lambda_0}(p^2)\phi(-p)-\phi'(-p)) \right]  \ ,
\label{functional integration kernel}
\end{align}
where $A_{\Lambda,\Lambda_0}$ and $B_{\Lambda,\Lambda_0}$ are defined
in (\ref{A}) and (\ref{B}), respectively, and satisfy 
the initial conditions
\begin{align}
A_{\Lambda_0,\Lambda_0} = \infty,  \;\;\; B_{\Lambda_0,\Lambda_0} = 1 \ .
\end{align}

Then, (\ref{exact renormalization group equation}) is solved in terms of ${\cal K}$ as
\begin{align}
e^{-S_{\Lambda}[\phi]}=\int {\cal D}\phi'  \ {\cal K}[\phi,\phi',\Lambda,\Lambda_0] \ 
e^{-S_{\Lambda_0}[\phi']} \ .
\label{solution to exact renormalization group equation}
\end{align}
By substituting (\ref{solution to exact renormalization group equation}) into 
the definition of $U$ (\ref{U}), performing $\phi$ integration, and 
using (\ref{explicit form of r})  and (\ref{explicit form of varphi}),
we can show (\ref{U=V}). Thus, we give another proof of the relation (\ref{relation}).

\section{$\epsilon$ expansion}
\setcounter{equation}{0}
In this section, as a check of the validity of the ERG equation in
(\ref{exact renormalization group equation}), we perform the $\epsilon$ expansion
in $d=4-\epsilon$ dimensions using the 
derivative expansion. We restrict ourselves to the case in which $\dot{C}_{\Lambda}(p^2)$ and 
$\chi_{\Lambda}(p^2)$ are given by (\ref{example of cdot}) and (\ref{example of chi}), respectively.
We will see that the scaling dimensions of operators around the Wilson-Fisher fixed point
are reproduced for arbitrary $\eta$ and $\zeta$.
The $\epsilon$ expansion using the derivative expansion was studied for the Polchinski equation
\cite{Polchinski:1983gv}, which can be viewed as an ERG equation, in \cite{ODwyer:2007brp} and for another specific ERG equation in \cite{Abe:2018zdc}. 
Here we use the procedure performed in\cite{Abe:2018zdc}.

We set $\Lambda_0=\infty$ so that 
(\ref{tau}) reduces to $\tau=1/\Lambda^2$.
We expand $\dot{C}_{\Lambda}$ in terms of $\tau$ as\footnote{Here we keep a factor
$e^{-\tau p^2}$ as a regularization of the delta function, which includes ambiguity.
Indeed, we have checked that the conclusion is unchanged if $e^{-\tau p^2}$ is replaced 
by $b e^{-\tau p^2}$ with $b$ being a constant.
A similar ambiguity in the regularization of the delta function is also seen
in (\ref{eq:|dphi/dvarphi|=}).}
\newpage
\begin{align}
\dot{C}_{\Lambda} (p^2) &= - e^{- \tau p^2} \Bigl \{ 2 \tau - \frac{\zeta}{p^2} \Bigl(1-e^{- \tau p^2} \Bigr) \Bigr \}  \notag \\
 &=-2 \tau e^{- \tau p^2} \Bigl(1- \frac{\zeta}{2} + O(\tau) \Bigr)   \ .
 \end{align}

Then, we rewrite (\ref{exact renormalization group equation}) in the coordinate space as
\begin{align} 
 \partial_{\tau}S_{\Lambda} [\phi] 
&= \int_x \Bigl(- \partial^2 \phi(x) - \frac{\eta}{4 \tau} \phi (x) \Bigr) \frac{\delta S_{\Lambda} [ \phi] }{\delta \phi (x)}  \notag \\
&\qquad - \frac{1}{2} \Bigl(1- \frac{\zeta}{2} \Bigr) \int_{x,y} K(x,y,\tau) \biggl \{ \frac{\delta S_{\Lambda}[\phi] }{ \delta \phi (x) } \frac{\delta S_{\Lambda}[\phi] }{ \delta \phi (y) } -\frac{\delta^2 S_{\Lambda}[\phi] }{ \delta \phi (x) \delta \phi (y)  }  \biggr \}   \ ,
\label{eq:FRG eq x space}
\end{align}
 where $\int_x= \int d^dx$ and $K(x,y,\tau)$ is defined in (\ref{heat kernel}).

 In the following, we perform the derivative expansion keeping up to two derivatives.
 The effective action is expanded as
 \begin{align}
 S_{\Lambda} [\phi] = \int_x \Bigl[ V_{\tau} (\phi(x)) + \frac{1}{2} W_{\tau} (\phi(x)) \bigl( \partial \phi(x) \bigr)^2 \Bigr] \ .   \label{eq:Stau derivative expansion}
\end{align}
By using the formula
\begin{align} 
 &\int_{x,y} K (x,y,\tau) f \bigl(\phi(x) \bigr) g \bigl(\phi(y) \bigr) \notag \\ 
 &= \int_x \biggl[ f \bigl(\phi(x) \bigr) g \bigl(\phi(x) \bigr) - \tau \bigl(\partial \phi(x) \bigr)^2 f' \bigl( \phi(x) \bigr) g' \bigl( \phi(x) \bigr) + O(\tau^2) \biggr]   \ , \label{eq:Fukuma Kfg=}
 \end{align}
we obtain, from (\ref{eq:FRG eq x space}), 
\begin{align} 
 &\partial_{\tau}S_{\Lambda} [\phi]  \notag \\
&= \int_x \biggl[ - \frac{\eta}{4 \tau} \phi V'_{\tau} - \frac{1}{2} \Bigl(1- \frac{\zeta}{2} \Bigr) V'^2_{\tau} + \frac{1}{2} \Bigl(1- \frac{\zeta}{2} \Bigr) (4 \pi \tau)^{-d/2} \Bigl(V''_{\tau} + \frac{d}{2\tau} W_{\tau} \Bigr)  \notag \\
&\qquad + \bigl( \partial \phi \bigr)^2 \Bigl \{ V''_{\tau} - \frac{\eta}{4 \tau} \Bigl( \frac{1}{2} \phi W'_{\tau} +W_{\tau}\Bigr)- \frac{1}{2} \Bigl(1- \frac{\zeta}{2} \Bigr) \Bigl(V'_{\tau} W'_{\tau} + 2V''_{\tau} W_{\tau} - \tau V''^2_{\tau} \Bigr)  \notag \\
&\qquad + \frac{1}{4} \Bigl(1- \frac{\zeta}{2} \Bigr) (4 \pi \tau)^{-d/2} W''_{\tau}   
 \Bigr \} \biggr]   \ .
 \label{derivative expansion 1}
 \end{align}
 On the other hand, from (\ref{eq:Stau derivative expansion}), we obtain
\begin{align}
  \partial_{\tau}S_{\Lambda} [\phi] = \int_x \Bigl[ \partial_{\tau}  V_{\tau} (\phi) + \frac{1}{2} \bigl( \partial \phi \bigr)^2 \partial_{\tau}  W_{\tau} (\phi) \Bigr]  \ .
\label{derivative expansion 2}
\end{align}
Comparing (\ref{derivative expansion 1}) and (\ref{derivative expansion 2}) gives rise to
\begin{align}
 \partial_{\tau} V_{\tau} (\phi) &= - \frac{\eta}{4 \tau} \phi V'_{\tau} - \frac{1}{2} \Bigl(1- \frac{\zeta}{2} \Bigr) V'^2_{\tau} + \frac{1}{2} \Bigl(1- \frac{\zeta}{2} \Bigr) (4 \pi \tau)^{-d/2} (V''_{\tau} + \frac{d}{2\tau} W_{\tau} ) \ ,  \label{deltau V} \\
 \partial_{\tau} W_{\tau} (\phi) &= 2V''_{\tau} - \frac{\eta}{2 \tau} \Bigl(\frac{1}{2} \phi W'_{\tau} +W_{\tau} \Bigr) - \Bigl(1- \frac{\zeta}{2} \Bigr) \Bigl(V'_{\tau} W'_{\tau} + 2V''_{\tau} W_{\tau} - \tau V''^2_{\tau} \Bigr) \notag \\
 &\qquad + \frac{1}{2}   \Bigl(1- \frac{\zeta}{2} \Bigr) (4 \pi \tau)^{-d/2} W''_{\tau}   \ .
  \label{eq:dtau W=2v''-}
\end{align}

We normalize the kinetic term in (\ref{eq:Stau derivative expansion}) as $W_{\tau} (\phi)=1$,
so that
\begin{align}
 W'_{\tau} (\phi) = 0, \qquad W''_{\tau} (\phi) =0  \label{eq:W'=0} \ .
\end{align}
Then, (\ref{deltau V}) and (\ref{eq:dtau W=2v''-}) reduce to
\begin{align}
 \partial_{\tau}V_{\tau} (\phi) &=- \frac{\eta}{4\tau} \phi V'_{\tau} - \frac{1}{2} \Bigl(1- \frac{\zeta}{2} \Bigr) V'^2_{\tau} + \frac{1}{2} \Bigl(1- \frac{\zeta}{2} \Bigr) (4 \pi \tau)^{-d/2} \Bigl(V''_{\tau} + \frac{d}{2\tau} \Bigr)  \ ,    \\
 \partial_{\tau}W_{\tau} (\phi) &=- \frac{\eta}{2\tau} + \zeta V''_{\tau} + \Bigl(1- \frac{\zeta}{2} \Bigr) \tau V''^2_{\tau}   \ . \label{eq:dtau W=tauV''2+}
\end{align}
We make a change of variable $\phi \to \varphi$ such that $W_{\tau+\delta\tau}=1$:
\begin{align}
 d\varphi (x) = \sqrt{W_{\tau + \delta\tau} (\phi) } \, d\phi (x)   \ .    \label{eq:varphi=sqrt W phi}
\end{align}
Then, the Jacobian for the change of variable is
\begin{align}
\biggl| \frac{\delta \phi}{\delta \varphi} \biggr| &= 1 - \delta\tau \int_{x,y} aK (x,y,\tau) \Bigl\{ - \frac{\eta}{4\tau} + \frac{\zeta}{2} V''_{\tau} +  \Bigl(1- \frac{\zeta}{2} \Bigr) \frac{ \tau}{2}  V''^2_{\tau}   \Bigr\}\delta^{(d)} (x-y)  \nonumber\\
&= \exp \Biggl[- \delta\tau\int_x a  (4 \pi \tau)^{-d/2} \biggl\{ - \frac{\eta}{4\tau} + \frac{\zeta}{2} V''_{\tau} + \Bigl(1- \frac{\zeta}{2} \Bigr) \frac{ \tau}{2}  V''^2_{\tau}   \biggr\} \Biggr]  \ ,
 \label{eq:|dphi/dvarphi|=}
\end{align}  
where we have regularized the delta function $\delta^{(d)}(x-y)$ by $aK(x,y,\tau)$, with $a$ a constant.

Further, we make quantities dimensionless as
\begin{align}
x \rightarrow \tau^{1/2}x, \;\;\; \varphi(x) \rightarrow \tau^{-(d-2)/4}\varphi(x), \;\;\;
V_{\tau} \rightarrow \tau^{-d/2}V_{\tau} \ .
\label{dimensionless}
\end{align}
Then, from (\ref{deltau V}), (\ref{eq:|dphi/dvarphi|=}) and (\ref{dimensionless}), we obtain
\begin{align}
\tau \partial_{\tau}  V_{\tau} = & \frac{d}{2} V_{\tau} -\frac{d-2}{4}\varphi V'_{\tau} - \frac{1}{2} \Bigl(1+ \frac{\zeta}{2} \Bigr)  V'^2_{\tau} + \frac{B}{2} \Bigl \{ 1 + \Bigl(a- \frac{1}{2} \Bigr) \zeta \Bigr \} V''_{\tau} + a \frac{B}{2} \Bigl(1- \frac{\zeta}{2} \Bigr) V''^2_{\tau}  \notag \\
&+ \frac{B}{4} \Bigl(d-a \eta - \frac{d}{2} \zeta \Bigr) - \frac{1}{2} \Bigl(1- \frac{\zeta}{2} \Bigr) V'_{\tau} \int^{\varphi}_0 d \varphi' V''^2_{\tau}   \ ,
 \label{eq:deltau V} 
\end{align}  
where $B=(4\pi)^{-d/2}$.

We expand $V_{\tau}$ in terms of $\varphi$ as
\begin{align}
V_{\tau} &= v_0 + \frac{1}{2!} v_2 \varphi^2 + \frac{1}{4!} v_4 \varphi^4 + \frac{1}{6!} v_6 \varphi^6 + \frac{1}{8!} v_8 \varphi^8 + \frac{1}{10!} v_{10} \varphi^{10} + \frac{1}{12!} v_{12} \varphi^{12} + \frac{1}{14!} v_{14} \varphi^{14} +\cdots  \ .
\label{eq:V= varphi expansion} 
\end{align} 
By substituting (\ref{eq:V= varphi expansion}) into (\ref{eq:deltau V}), we obtain
\begin{align}
\tau \partial_{\tau} v_2 = &v_2 - \Bigl(1+ \frac{\zeta}{2} \Bigr)v_2^2- \Bigl(1- \frac{\zeta}{2} \Bigr)v^3_2+ \frac{B}{2} \Bigl \{1+ \Bigl(a- \frac{1}{2} \Bigr) \zeta \Bigr \}v_4  \notag \\
&+ aB \Bigl(1- \frac{\zeta}{2} \Bigr)v_2v_4 \  ,   \label{eq:flow eq v2 g=0}   \\
\tau \partial_{\tau} v_4 = & \frac{\epsilon}{2} v_4 - 4 \Bigl(1+ \frac{\zeta}{2} \Bigr)v_2 v_4 -6 \Bigl(1- \frac{\zeta}{2} \Bigr)v^2_2 v_4+ \frac{B}{2} \Bigl \{ 1+ \Bigl(a- \frac{1}{2} \Bigr) \zeta \Bigr \} v_6  \notag \\
& +aB \Bigl(1- \frac{\zeta}{2} \Bigr)(3v^2_4+v_2v_6) \  ,   \label{eq:flow eq v4 g=0}  \\
\tau \partial_{\tau} v_6 = &(-1+ \epsilon) v_6 - \Bigl(1+ \frac{\zeta}{2} \Bigr) (10v_4^2 + 6v_2 v_6) + \frac{B}{2} \Bigl \{ 1+ \Bigl(a- \frac{1}{2} \Bigr) \zeta \Bigr \} v_8  \notag \\
&+aB \Bigl(1- \frac{\zeta}{2} \Bigr)(15v_4 v_6+v_2 v_8)- \Bigl(1- \frac{\zeta}{2} \Bigr)(38v_2 v^2_4+ 9v^2_2 v_6) \ ,   \label{eq:flow eq v6 g=0}  \\
\tau \partial_{\tau} v_8 = & \Bigl(-2+ \frac{3}{2} \epsilon \Bigr) v_8 - \Bigl(1+ \frac{\zeta}{2} \Bigr)(56v_4 v_6+8v_2 v_8) + \frac{B}{2} \Bigl \{ 1+ \Bigl(a- \frac{1}{2} \Bigr) \zeta \Bigr \}v_{10}  \notag \\
&+aB \Bigl(1- \frac{\zeta}{2} \Bigr)(35v^2_6+28v_4 v_8+v_2 v_{10} )   \notag \\
&- \Bigl(1- \frac{\zeta}{2} \Bigr)(12v^2_2 v_8+168v^3_4+232v_2 v_4 v_6) \  ,  \\
\tau \partial_{\tau} v_{10} = &(-3+2 \epsilon) v_{10} - \Bigl(1+ \frac{\zeta}{2} \Bigr)(126v_6^2+120v_4 v_8+10v_2 v_{10} )  \notag \\
&+ \frac{B}{2} \Bigl \{ 1+ \Bigl(a- \frac{1}{2} \Bigr) \zeta \Bigr \} v_{12}+aB \Bigl(1- \frac{\zeta}{2} \Bigr)(210v_6 v_8+45v_4 v_{10}+v_2 v_{12})   \notag \\
&- \Bigl(1- \frac{\zeta}{2} \Bigr)(602v_2 v^2_6+520v_2 v_4 v_8+15v^2_2 v_{10}+2556v^2_4 v_6) \ ,  \label{eq:flow eq v10 g=0} \\
\tau \partial_{\tau} v_{12} = & \Bigl(-4+ \frac{5}{2} \epsilon \Bigr) v_{12} - \Bigl(1+ \frac{\zeta}{2} \Bigr)(792v_6 v_8+220v_4 v_{10}+12v_2 v_{12} )  \notag \\
&+ \frac{B}{2} \Bigl \{ 1+ \Bigl(a- \frac{1}{2} \Bigr) \zeta \Bigr \} v_{14}+aB \Bigl(1- \frac{\zeta}{2} \Bigr)(462v_8^2+495v_6 v_{10}+66v_4 v_{12}+v_2v_{14})   \notag \\
&- \Bigl(1- \frac{\zeta}{2} \Bigr)(4104v_2 v_6v_8+980v_2 v_4 v_{10}+18v^2_2 v_{12}+19580v_4v_6^2+8536v_4^2v_8)  \ ,  \label{eq:flow eq v12 g=0} 
 \end{align} 
 where $\epsilon = 4-d$.

We look for the Wilson-Fisher fixed point by assuming that
\begin{align}
v_2^*=O(\epsilon) , \qquad v_4^*=O(\epsilon) , \qquad v_6^*=O(\epsilon^2) , \qquad v_n^*=O(\epsilon^3) 
\;\; \mbox{for} \;\; n \geq 8   \ .\label{eq:Wilson-Fisher fixed point}
\end{align}
We set $\tau \partial_{\tau} v_n^* =0$ in 
(\ref{eq:flow eq v2 g=0})-(\ref{eq:flow eq v6 g=0}), and expand the right-hand side of (\ref{eq:flow eq v2 g=0})  up
to the first order in $\epsilon$ 
and those of
(\ref{eq:flow eq v4 g=0}) and (\ref{eq:flow eq v6 g=0}) up to the second order in $\epsilon$.
Then, we obtain equations determining the fixed point as follows:
\begin{align}
0 &= v_2^* + \frac{B}{2} \Bigl \{ 1+ \Bigl(a- \frac{1}{2} \Bigr) \zeta \Bigr \} v_4^* \ , \label{eq:0=v2*+v4* g=0} \\
0 &= \frac{\epsilon}{2} v_4^* - 4 \Bigl(1+ \frac{\zeta}{2} \Bigr) v_2^* v_4^* +3aB \Bigl(1- \frac{\zeta}{2} \Bigr) (v_4^*)^2 + \frac{B}{2} \Bigl \{ 1+ \Bigl(a- \frac{1}{2} \Bigr) \zeta \Bigr \} v_6^* \  , \\
0 &= -v_6^* -10 \Bigl(1+ \frac{\zeta}{2} \Bigr) (v_4^*)^2 \ . \label{eq:0=v6*+v4* g=0} 
\end{align}
We solve these equations as
\begin{align}
v_2^* &= - \frac{\epsilon}{6A} \Bigl \{ 1+ \Bigl(a- \frac{1}{2} \Bigr) \zeta \Bigr \} +O(\epsilon^2)  \ , \label{eq:v2*=-e/6A} \\ 
v_4^* &= \frac{\epsilon}{3B_0A} +O(\epsilon^2)   \ ,   \label{eq:v4*=e/3BA} \\
v_6^* &= - \frac{10 \epsilon^2}{(3B_0A)^2} \Bigl(1+ \frac{\zeta}{2} \Bigr)+O(\epsilon^3)  \ , \label{eq:v6*=-10e2/3BA2}  
\end{align}
where
\begin{align}
A \equiv \Bigl(a- \frac{1}{2} \Bigr) \zeta^2 +3a \zeta -2a+2  \label{eq:A eqiv (a-1/2)} \ .
\end{align}
We take $a$ such that $A \neq 0$.
Note that
\begin{align}
B=(4 \pi)^{-d/2}=B_0+ O(\epsilon) = (4\pi)^{-2}+O(\epsilon)  \ .
\end{align}

By substituting $v_n=v_n^*+ \delta v_n$ into (\ref{eq:flow eq v2 g=0})-(\ref{eq:flow eq v12 g=0})
and expanding the right-hand side of (\ref{eq:flow eq v2 g=0})-(\ref{eq:flow eq v12 g=0}) up to the
first order in $ \epsilon$ and $\delta v_n$,
we obtain
\begin{align}
\tau \partial_{\tau} \delta v_2 =& \Bigl \{1-2 \Bigl(1+ \frac{\zeta}{2} \Bigr) v_2^*+aB \Bigl(1- \frac{\zeta}{2} \Bigr)  v_4^* \Bigr \} \delta v_2  \notag \\
&+ \frac{B}{2} \Bigl \{1+ \Bigl(a- \frac{1}{2} \Bigr) \zeta +2a \Bigl(1- \frac{\zeta}{2} \Bigr) v_2^* \Bigr \} \delta v_4 \  , \label{eq:flow eq v2 g=0 v*} \\
\tau \partial_{\tau} \delta v_4 =&-4 \Bigl(1+ \frac{\zeta}{2} \Bigr) v_4^* \delta v_2 + \Bigl \{ \frac{\epsilon}{2} -4 \Bigl(1+ \frac{\zeta}{2} \Bigr) v_2^* +6aB \Bigl(1- \frac{\zeta}{2} \Bigr) v_4^* \Bigr \} \delta v_4   \notag \\
&+ \frac{B}{2} \Bigl \{1+ \Bigl(a- \frac{1}{2} \Bigr) \zeta +2a \Bigl(1- \frac{\zeta}{2} \Bigr) v_2^* \Bigr \} \delta v_6  \  , \\
\tau \partial_{\tau} \delta v_6 =&-20 \Bigl(1+ \frac{\zeta}{2} \Bigr)v_4^* \delta v_4 + \Bigl \{-1+ \epsilon -6 \Bigl(1+ \frac{\zeta}{2} \Bigr)v_2^*+15aB \Bigl(1- \frac{\zeta}{2} \Bigr)v_4^* \Bigr \} \delta v_6   \notag \\
&+ \frac{B}{2} \Bigl \{ 1+ \Bigl(a- \frac{1}{2} \Bigr) \zeta +2a \Bigl(1- \frac{\zeta}{2} \Bigr)v_2^* \Bigr \} \delta v_8 \  , \\
\tau \partial_{\tau} \delta v_8 =&-56 \Bigl(1+ \frac{\zeta}{2} \Bigr)v_4^* \delta v_6 + \Bigl \{-2+ \frac{3}{2} \epsilon -8 \Bigl(1+ \frac{\zeta}{2} \Bigr)v_2^*+28aB \Bigl(1- \frac{\zeta}{2} \Bigr)v_4^* \Bigr \} \delta v_8  \notag \\
&+ \frac{B}{2} \Bigl \{1+ \Bigl(a- \frac{1}{2} \Bigr) \zeta+2a \Bigl(1- \frac{\zeta}{2} \Bigr)v_2^* \Bigr \} \delta v_{10} \  ,\\
\tau \partial_{\tau} \delta v_{10} =&-120 \Bigl(1+ \frac{\zeta}{2} \Bigr)v_4^* \delta v_8 + \Bigl \{-3 + 2 \epsilon - 10 \Bigl(1+ \frac{\zeta}{2} \Bigr)v_2^* +45aB \Bigl(1- \frac{\zeta}{2} \Bigr)v_4^* \Bigr \} \delta v_{10}   \notag \\
&+ \frac{B}{2} \Bigl \{ 1+ \Bigl(a- \frac{1}{2} \Bigr) \zeta +2a \Bigl(1- \frac{\zeta}{2} \Bigr) v_2^* \Bigr \} \delta v_{12} \  , \label{eq:flow eq v10 v*}  \\
\tau \partial_{\tau} \delta v_{12} =&-220 \Bigl(1+ \frac{\zeta}{2} \Bigr)v_4^* \delta v_{10} + \Bigl \{ -4 + \frac{5}{2} \epsilon - 12 \Bigl(1+ \frac{\zeta}{2} \Bigr)v_2^* +66aB \Bigl(1- \frac{\zeta}{2} \Bigr)v_4^* \Bigr \} \delta v_{12}   \notag \\
&+ \frac{B}{2} \Bigl \{ 1+ \Bigl(a- \frac{1}{2} \Bigr) \zeta +2a \Bigl(1- \frac{\zeta}{2} \Bigr) v_2^* \Bigr \} \delta v_{14}  \ . \label{eq:flow eq v12 v*}
 \end{align}    
We regard these as a linear transformation for $\delta v_n$.
Then, the eigenvalue equation reads
\begin{align}
&- \lambda^5+ \lambda^4 \Bigl \{-5 +5 \epsilon -15(2+ \zeta)v_2^* + \frac{95}{2} aB(2- \zeta) v_4^* \Bigr \}   \notag \\
&+ \lambda^3 \Bigl \{ -5 +15 \epsilon -50(2+ \zeta) v_2^* +B(-100+270a-235a \zeta +25 \zeta^2 -50a \zeta^2)v_4^* \Bigr \}  \notag \\
&+ \lambda^2 \Bigl \{ 5 -15(2+ \zeta)v_2^* +B(-108-17a- \frac{199}{2}a \zeta +27 \zeta^2 -54a \zeta^2)v_4^* \Bigr \}   \notag \\
&+ \lambda \Bigl \{ 6 -17 \epsilon +44(2+ \zeta)v_2^* +B(112-288a+256a \zeta -28 \zeta^2 +56a \zeta^2)v_4^* \Bigr \} \notag \\
&-3 \epsilon +12(2+ \zeta)v_2^* +B(48-36a+66a \zeta -12 \zeta^2 +24a \zeta^2)v_4^* =0  \ . \label{eq:lambda5+lambda4+=0 g=0}
\end{align}
Here we have ignored  (\ref{eq:flow eq v12 v*}) and dropped the term proportional to
$\delta v_{12}$ in (\ref{eq:flow eq v10 v*}).
By substituting (\ref{eq:v2*=-e/6A}) and (\ref{eq:v4*=e/3BA}) into this equation, we obtain
\begin{align}
- \lambda^5 &- \lambda^4 \Bigl \{5 - \frac{\epsilon}{6} (45+110D)  \Bigr \} - \lambda^3 \Bigl \{5 - \frac{\epsilon}{3} (20+110D) \Bigr \}   \notag \\
&+ \lambda^2 \Bigl \{ 5 - \frac{\epsilon}{6} (93+110D) \Bigr \} + \lambda \Bigl \{ 6 - \frac{\epsilon}{3} (17+110D) \Bigr \} +3 \epsilon =0  \ , \label{eq:lambda5+lambda4+=0 g=0 e}
\end{align}
where
\begin{align}
D \equiv \frac{a(2- \zeta)}{A}= \frac{a(2- \zeta)}{(a-1/2) \zeta^2 +3a \zeta -2a+2}  \ . \label{eq:D=a()/A}
\end{align}
The solutions to (\ref{eq:lambda5+lambda4+=0 g=0 e}) are
\begin{align}
\lambda_2 &= 1- \frac{1}{6} \epsilon, \quad \lambda_4= - \frac{1}{2} \epsilon, \quad \lambda_6=-1- \frac{3}{2} \epsilon, \quad \lambda_8=-2- \frac{19}{6} \epsilon,  \notag \\
\lambda_{10} &=-3+ \frac{\epsilon}{6} (77+110D)  \ .    \label{eq:eigen value 5 g=0}
\end{align}
$\lambda_2 \sim \lambda_8$ are the well-known scaling dimensions of operators 
around the Wilson-Fisher fixed point.
We have checked 
that $\lambda_{10}$ is also obtained correctly if  (\ref{eq:flow eq v12 v*}) is included.

\section{Conclusion and discussion}
\setcounter{equation}{0}
We have studied the relationship between the renormalization group 
and the diffusion equation. We considered the
ERG equation for a scalar field that includes an arbitrary cutoff function and
an arbitrary quadratic seed action. As a generalization of the result in ref.\cite{Sonoda:2019ibh}, 
we found that the correlation functions of diffused fields with respect to the bare action
agree with those of bare fields with respect to the effective action,
where the diffused field obeys a generalized diffusion equation determined by the cutoff function
and the seed action and  agrees 
with the bare field at the initial time.
This result is reasonable in that diffusion is associated with coarse-graining and 
the coarse-graining procedure is fixed by
the cutoff function and the seed action.
We performed the $\epsilon$ expansion using the derivative expansion
as a check of the validity of the ERG equation.
We reproduced the well-known scaling dimensions of operators around the Wilson-Fisher fixed point.

We comment on 
a case in which the seed action includes terms that are higher than the second order in $\phi$.
In this case, $U$ satisfies a functional differential equation given by the fourth equality in 
(\ref{calculation for U}), while $V$ satisfies a functional differential equation given by the third 
equality in (\ref{calculation for V}). The latter equation includes terms that mix 
$\delta V/\delta J$ with $\delta R/\delta J$, and those terms do not exist in the former equation.
This implies that the relation (\ref{relation})  does not hold for this case even if $r$ is modified.
We need to generalize (\ref{relation}) in some way.

Finally, we note that (\ref{general diffusion equation}) is a "gradient flow equation",  
\begin{align}
\partial_{\tau}\varphi(\tau,p) = -\frac{\delta \tilde{S}}{\delta \varphi(\tau,-p)}
\end{align}
with
\begin{align}
\tilde{S}=\int_p \frac{\Lambda^2}{4} \chi_{\Lambda}(p^2) \varphi(\tau,p)\varphi(\tau,-p) \ .
\end{align}
While the above gradient flow equation is at first sight conceptually different from the one
studied in the context of gauge theories, we hope that our findings in this paper will
give some insights into construction of an ERG equation with manifest
gauge invariance.

\section*{Acknowledgments}
A.T.\ was supported in part by Grant-in-Aid for Scientific Research
(No. 18K03614) from the Japan Society for the Promotion of Science.

\end{document}